\documentclass[aps,prl,twocolumn,groupedaddress]{revtex4}
\usepackage{amsbsy,amssymb,amsmath,bm} \usepackage{graphicx}

\input{epsf}

\begin{document}

\title{Localized Superconductivity in the Quantum-Critical Region of the
Disorder-Driven Superconductor-Insulator Transition in TiN Thin Films
}

\author{T.\,I.~Baturina$^{1,2}$\/\email{tatbat@isp.nsc.ru}, A.\,Yu. Mironov$^{1,2}$,
         V.\,M.~Vinokur$^3$, M.\,R.~Baklanov$^4$,  and C.~Strunk$^2$}
\affiliation{
$^1$Institute\,of\,Semiconductor\,Physics,\,630090,\,Novosibirsk,\,Russia\\
$^2$Institut\,f\"{u}r\,experimentelle\,und\,angewandte\,Physik,\,Universit\"{a}t\,Regensburg,\,D-93025\,Regensburg,\,Germany\\
$^3$Materials\,Science\,Division,\,Argonne\,National\,Laboratory,\,Argonne,\,IL\,60439,\,USA\\
$^4$IMEC, Kapeldreef 75, B-3001 Leuven, Belgium}

\date{\today}

\begin{abstract}
We investigate low-temperature transport properties of thin TiN
superconducting films in the vicinity of the disorder-driven
superconductor-insulator transition.  In a zero magnetic field, we
find an extremely sharp separation between superconducting and
insulating phases, evidencing a direct superconductor-insulator
transition without an intermediate metallic phase.  At moderate
temperatures, in the insulating films we reveal thermally activated
conductivity with the magnetic field-dependent activation energy. At
very low temperatures, we observe a zero-conductivity state, which
is destroyed at some depinning threshold voltage $V_T$. These
findings indicate formation of a distinct collective state of the
localized Cooper pairs in the critical region at both sides of the
transition.
\end{abstract}


\maketitle

An early suggestion that tuning disorder strength can cause a direct
superconductor-insulator transition (SIT) in two-dimensional
systems~\cite{Fisher89} triggered explosive activity in experimental
studies of superconductor films~\cite{Goldman_Review}.
Experimentally, the SIT can be induced by decreasing the film
thickness \cite{GoldmanBi} and/or, close to the critical thickness,
also by the magnetic field~\cite{Hebard}. These scenarios are
commonly referred to as disorder-driven SIT (D-SIT) and
magnetic-field driven SIT. Recent studies on the $B$-induced
insulator revealed several striking features: 
a magnetic-field-dependent thermally activated behavior of the
conductivity~\cite{Shahar04} and a threshold response to the dc
voltage~\cite{Shahar05}, indicating the possible formation of a distinct
collective insulating state. Importantly, these findings refer to
the films belonging to the superconducting side of the D-SIT. This
rises the question of whether the above findings are specific only to
the superconducting side of the D-SIT or a characteristic feature of
the whole critical region including both the insulating and
superconducting sides of the D-SIT.

In this Letter we focus on the insulating side of the
disorder-driven superconductor-insulator transition in TiN films.
The transition itself turns out to be exceptionally sharp. At zero
and low magnetic fields we find thermally activated behavior of the
conductivity. A positive magnetoresistance and a distinct threshold
behavior in the low-temperature $I$-$V$ characteristics persist on the
{\it insulating} side of the D-SIT. Our results clearly indicate
that, in the vicinity of the D-SIT, the response to applied magnetic
and/or electric fields, is the same irrespective of whether the
underlying ground state is superconducting or insulating.

The 5-nm thick TiN films were grown by atomic layer chemical vapor
deposition onto a Si/SiO$_2$ substrate. The samples for transport
measurements were patterned into Hall bridges using conventional UV
lithography and subsequent plasma etching. To increase sheet
resistances ($R_\square$) without introducing structural changes,
the films were thinned by an additional soft plasma etching.
Electron transmission micrographs and diffraction patterns revealed
a polycrystalline structure in both initial and etched films, the
interfaces separating densely-packed crystallites being 1--2 atomic
layers thick. As we found before~\cite{OurTiN04}, in such samples
$k_F\ell\simeq 1$, where $k_F$ is the Fermi vector, and $\ell\sim
0.3$\,nm is the mean free path. This short mean free path can be related
to the enhanced Cl content (up to $3\%$), characteristic of films
grown by the above method~\cite{SattaTiN}. Four-probe resistance
measurements were carried out by the standard low frequency 
(0.4--2\,Hz) ac lock-in current source technique with the ac current 
0.01--1\,nA. In cases where the resistances were too high to employ the
current source four-probe lock-in measurements, the two-probe
voltage source technique with the ac voltage 10--30\,$\mu$V was
used.  At resistances $R>1$\,M$\Omega$, this allowed us to keep the
power dissipation below $10^{-15}$\,W, thus ensuring linear regime
and excluding overheating. Two-probe differential conductance vs dc
voltage measurements were done by means of the low frequency ac
lock-in technique combined with the dc voltage excitations. Magnetic
fields up to 16\,T were applied perpendicular to the film surface.

\begin{figure}[tb]
\includegraphics[width=80mm]{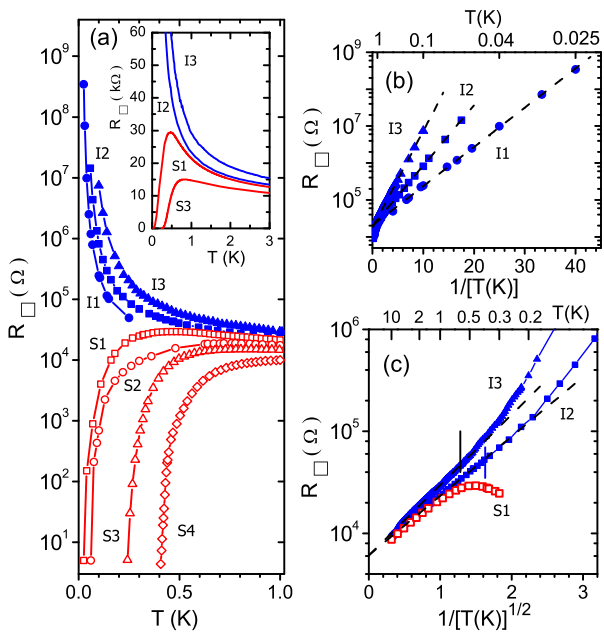}
\caption{\label{fig1:RTB0fan} (color online). Temperature dependences of 
$R_\square$ taken at zero magnetic field for the samples near the
localization threshold. (a)~$\log R_\square$ versus $T$. Inset: some
part of the $R_\square$ data in a linear scale. (b)~$\log R_\square$
versus $1/T$ for samples I1, I2, and I3. Dashed lines represent
Eq.~(1) and fit perfectly at low temperatures. All curves saturate
at the same  $R_\square\approx 20\,{\rm k}\Omega$ at high temperatures. 
(c)~$R_\square$ versus
$1/T^{1/2}$; dashed lines are given by $R_\square =R_1
\exp(T_1/T)^{1/2}$ which (with $R_1 \sim 6\,{\rm k}\Omega$) well fit
the data at high temperatures. Vertical strokes mark $T_0$,
determined by the fit to the Arrhenius formula of Eq.~(1).}
\end{figure}

We start with the zero magnetic-field results.
Figure~\ref{fig1:RTB0fan}(a) shows the temperature dependence of
$\log R_\square$ for seven different films. An increase in disorder
results in a growth of $R_\square$ and reduces the superconductor
critical temperature monotonically. We did not detect any sign of
the reentrant behavior or a kink in $R(T)$, which are characteristic
to the granular films and/or the films containing large scale
inhomogeneities~\cite{granular}. It indicates that our films do not
have a granular structure but are rather homogeneously disordered.
To characterize the behavior on the nonsuperconducting side, we
replot the data for $R_\square$ versus $1/T$  in
Fig.~\ref{fig1:RTB0fan}(b). At low temperatures we observe an
Arrhenius behavior of the resistance, demonstrating that these
samples are indeed insulators. The dashed lines correspond to
\begin{equation}
R =R_0 \exp(T_0/T).
\label{main}
\end{equation}
The activation temperatures $T_0$ in the three samples measured are
$T_0 = 0.25$\,K (I1), $0.38$\,K (I2), and $0.61$\,K (I3) (the growth
in $T_0$ corresponds to increasing disorder). The prefactor $R_0$
determined from the extrapolation of the dashed lines in
Fig.\,\ref{fig1:RTB0fan}(a) towards $1/T=0$ is almost the same
($\approx 20\,{\rm k}\Omega$) for all samples.
Figure\,\ref{fig1:RTB0fan} shows that our TiN films demonstrate an
abrupt switch between the superconducting and insulating phases:
indeed, the ``last'' superconducting and the ``first'' insulating films
are practically indistinguishable by their resistances at
temperatures higher than 1\,K, for instance, at $T=10\,{\rm K}$
$R_\square=8.74\,{\rm k}\Omega$ (S1) and $R_\square=9.16\,{\rm
k}\Omega$ (I2). However, at lower temperatures they choose
unequivocally between either the superconducting or insulating
ground states.

\begin{figure}[tb]
\includegraphics[width=70mm]{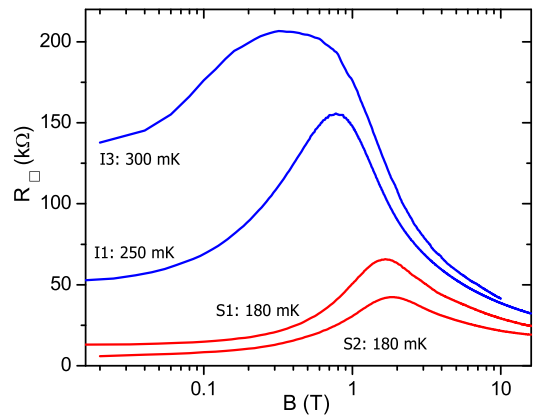}
\caption{\label{fig2:RBhigh} (color online). Magnetoresistance
isotherms for superconducting (S1, S2) and insulating samples (I1,
I3) at similar temperatures. All curves converge above 2\,T. }
\end{figure}

We point out that the separatrix between the sets of
``superconducting'' and ``insulating'' $R(T)$ curves is not simply a
horizontal line [$R(T)={\rm const}$]. To demonstrate that, we replot
the data in the linear scale in the inset to
Fig.~\ref{fig1:RTB0fan}(a) and note that $R(T)$ dependences of
superconducting samples are nonmonotonic. They exhibit an
``insulating trend," i.e. an upward turn of the resistance preceding
its eventual drop to zero at low temperatures.  Further insight into
the evolution of TiN films across the D-SIT is drawn from the $\log
R_{\square}$ against $1/T^{1/2}$ plots shown in
Fig.~\ref{fig1:RTB0fan}(c). At $T>T_0$, the resistances of the
insulating samples compare favorably with the Efros-Shklovskii (ES)
formula, $R =R_1 \exp(T_1/T)^{1/2}$,~\cite{Shklov,ES}. The
temperatures $T_1$ in the three samples shown in
Fig.~\ref{fig1:RTB0fan}(c) are $T_1 = 1.75$\,K (S1), $1.80$\,K
(I2), and $2.53$\,K (I3). The prefactor $R_1$ is again nearly the
same for all samples, but in this case it is close to the quantum
resistance \textit{for pairs}, $h/(2e)^2=6.45\,{\rm k}\Omega$. At
lower temperatures $R(T)$ deviates from the ES behavior, which in
the insulating samples turns into the Arrhenius law below $T_0$,
while the superconducting samples fall into a superconducting state.

\begin{figure}[tb]
\includegraphics[width=70mm]{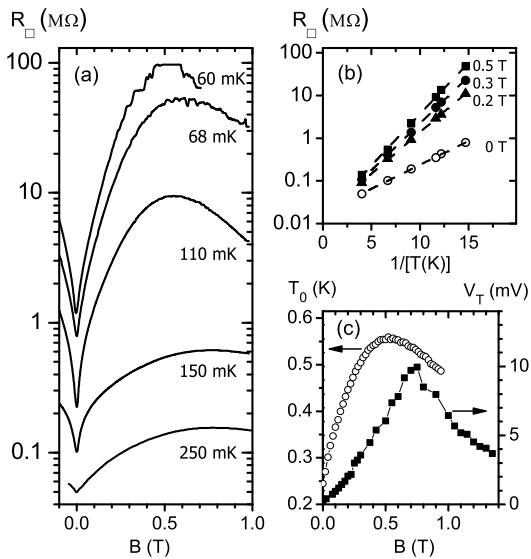}
\caption{\label{fig3:RBlow} (a) Sheet resistance of sample I1 as a
function of the magnetic field at some temperatures listed. (b) $R$
versus $1/T$ at $B=0$ (open circles), 0.2 (triangles), 0.3 (filled
circles), and 0.5\,T (squares). The dashed lines are given by
Eq.\,(1). (c) $T_0$ (left axis), calculated from fits to Eq.\,(1),
and the threshold voltage $V_T$ (right axis) as a function of $B$. }
\end{figure}

Turning to the magnetoresistance data shown in
Fig.~\ref{fig2:RBhigh}, we see that in all samples, including the
insulating films, $R(B)$ varies nonmonotonically with $B$. It
exhibits a positive magnetoresistance (PMR) at low fields, then
reaches a maximum, followed first by a rapid drop, and eventually
saturates at higher magnetic fields \cite{batPRL07}, where the
difference between insulating and superconducting samples is
suppressed and all curves converge. Since PMR in superconducting
films appears because of the suppression of superconducting phase
coherence by the magnetic field, one can conjecture that this phase
coherence persists also in \textit{insulating} films. At low
temperature the ratio of the magnitude of the resistance at maximum
to its value at zero magnetic field increases. Figure~\ref{fig3:RBlow}
presents $R(B)$ of the sample I1 at low temperatures and shows that
the resistance at low fields again follows the Arrhenius behavior
with $B$-dependent $T_0$ [Eq.(1)]. The activation temperature
$T_0(B)$ qualitatively follows the magnetoresistance.
\begin{figure}[tb]
\includegraphics[width=70mm]{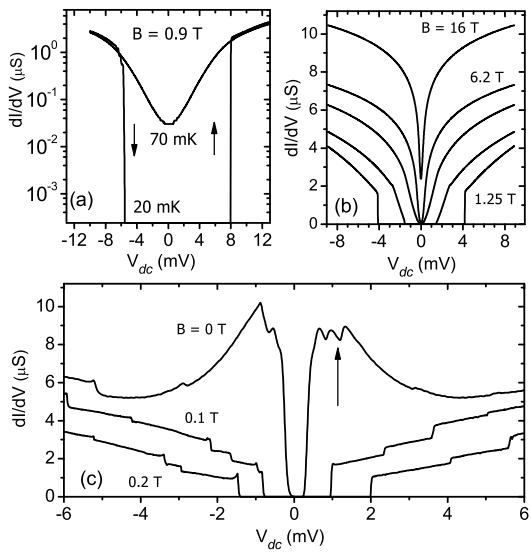}
\caption{\label{fig4:VAch} Differential conductance vs dc voltage
for sample I1. (a)~$dI/dV(V_{dc})$ at $B=0.9$\,T at two temperatures
$T=0.02$ and 0.07\,K. An arrow shows the direction of the voltage
sweep. (b)~$dI/dV(V_{dc})$ at $B=1.25$, 1.65, 2.7, 6.2, and 16\,T at
$T=0.02$\,K. (c)~$dI/dV(V_{dc})$ at $B=0$, 0.1, and 0.2\,T at
$T=0.02$\,K.}
\end{figure}

We now discuss the most intriguing feature, a depinning transition
observed at low temperatures in the insulating films.
Figure~\ref{fig4:VAch} reveals an abrupt onset of finite
conductivity as the bias voltage $V_{dc}$ exceeds a threshold
voltage $V_T$: panel (a) presents a differential conductance,
$dI/dV$ versus $V_{dc}$ measured on sample I1 at $B=0.9$\,T, using
the two-probe technique with the contact separation $1.5$\,mm. Two
traces represent data taken at $T=70$\,mK and at the lowest
temperature, $T=20$\,mK, achieved in the experiment. The
$dI/dV$ curve at $T=70$\,mK  is typical for an insulator with
activated conductance, showing a gradual increase of $dI/dV$ with
$V_{dc}$, and is symmetric with respect to the direction of the
$V_{dc}$ sweep. However, as temperature is decreased to $T=20$~mK
the voltage response changes dramatically. Under low $V_{dc}$, both,
the current and $dI/dV$ are immeasurably small. As soon as $V_{dc}$
reaches some well-defined threshold value $V_T$, $dI/dV$ abruptly
jumps up by several orders of magnitude. The threshold behavior is
accompanied by a hysteresis. A sharp conductance jump is observed up
to $B\sim 2$\,T. A non-Ohmic behavior remains even at $B=16$~T. The
threshold voltage changes nonmonotonically upon magnetic field [see
$V_T(B)$ along with $T_0(B)$ in Fig.~\ref{fig3:RBlow}(c)]. Note the
large magnitude of threshold voltage to activation energy ratio:
$eV_T/k_BT_0\approx 220$ at $B=0.7$\,T.

Notable is also the nonmonotonic behavior of $dI/dV(V_{dc})$ at
$B=0$, which is displayed in Fig.~\ref{fig4:VAch}(c). Similar to the
depinning-like behavior at finite magnetic field in
Figs.~\ref{fig4:VAch}(a) and (b), we find a steep initial increase of
$dI/dV(V)$ with a maximum around $V_{dc}=1$~mV (marked by an arrow),
followed by a gradual decrease. At the peak, $dI/dV$ is about twice
larger than at $V_{dc}=4$~mV. This feature cannot be explained by
electron heating and vanishes already at $B=0.025$~T. Such a
suppression of the conductivity by the bias current or voltage is
typical for a superconductor in a fluctuation regime. The
observation a similar ``superconducting trend'' in the insulating
regime indicates that superconducting correlations survive also on
the insulating side of the D-SIT. This implies the presence of the
localized Cooper pairs and local superconducting phase coherence.

To summarize the essential findings in the D-SIT critical region, we
note that the search for a disorder-driven superconductor-insulator
transition has included many materials, e.g., Bi~\cite{GoldmanBi},
MoSi~\cite{MoSiOkuma}, Ta~\cite{TaYoon},
InO$_x$~\cite{ShaharOvadyahu,VFGInOIns,Shahar04}, and Be~\cite{Be1}.
However, the immediate onset of exponential temperature dependence
of the resistance, which conclusively evidences the direct
transition into an insulator, was found so far only in
InO$_x$~\cite{ShaharOvadyahu,VFGInOIns,Shahar04} and Be~\cite{Be1}
films. In Bi, MoSi, and Ta 
compounds~\cite{GoldmanBi,MoSiOkuma,TaYoon} a weak logarithmic
temperature dependence of the resistance was observed on the
nonsuperconducting side, either because of a possible intermediate
metallic phase or since the transition to the superconducting state
occurs there at much lower temperatures. Our data on TiN
unambiguously show a direct D-SIT with the
\textit{nonhorizontal} $R(T)$ separatrix between the insulating and
superconducting sides.  An insulating trend (i.e. the upturn of the
separatrix) can also be seen in the data on
InO$_x$~\cite{ShaharOvadyahu} and Be~\cite{WuES} films. This implies
that D-SIT \textit{cannot} be described by a single-parameter
scaling with the universal resistance at the transition.

The next important feature is that in all three materials exhibiting
D-SIT, TiN, InO$_x$, and Be,~\cite{KowalOvadyahu,ShaharOvadyahu,Shahar04,Be1}, the
low-temperature activated behavior, $R(T)\propto\exp(T_0/T)$
transforms into a variable range hopping upon increasing
temperature, contrary to common wisdom expectations. The fact that
the high-temperature behavior of the last superconducting sample,
S1, is close to that of I2 [$R(T)$s are nearly indistinguishable at
high temperatures; see Fig.\,1] evidences the insulating features of
superconducting samples in the critical region. On the other hand,
the similarity in $R(T)$ between superconducting and insulating
samples in the ES regime indicates the presence of Cooper pairs in
the insulating samples as well. Further, all these materials exhibit
positive magnetoresistance on the insulating side of the D-SIT
(InO$_x$~\cite{VFGInOIns} and Be films~\cite{BeInsPMR,Butko}). And,
finally, nonmonotonic dependencies of $T_0(B)$, extracted from the
Arrhenius behavior of $R(T)$, and those of the threshold voltage,
$V_T(B)$, which we find at the insulating side of D-SIT, have also
been observed in the magnetic-field-induced insulating phase in  samples,
which are superconductors at zero $B$~\cite{Shahar04,Shahar05}.

We would like to emphasize that the investigated films are
homogeneously disordered and do not possess structural granularity.
A positive magnetoresistance and thresh\-old behavior, recently
reported by~\cite{GoldmanGr}, were observed in \textit{structurally
granular} systems at temperatures where granules are
superconducting, while disappearing with the breakdown of
superconductivity in granules.

From all the above we conclude that in the critical region of the
transition a peculiar highly inhomogeneous insulating phase with
superconducting correlations, a \textit{Cooper-pair insulator},
forms. In other words, the films in the critical region may be
regarded as possessing a self-induced granularity due to strong
mesoscopic fluctuations in disordered superconducting thin
films~\cite{feigelman05} giving rise to formation of superconducting
droplets (islands) immersed into an insulating matrix.  Such a state
can be viewed as a network of superconducting islands coupled by
weak links (Josephson junctions arrays).  This is strongly supported
by experiments of~\cite{Haviland}, where the voltage threshold
similar to ours was observed on the long chains of SQUIDs.

The Cooper-pair insulator establishes as a result of the mutual
Josephson phase locking and exhibits generic collective
behavior~\cite{FVB}. At moderate temperatures, it shows thermally
activated conductivity governed by the large collective Coulomb
blockade gap for a Cooper-pair propagation.  The observed voltage
threshold behavior suggests that the Cooper-pair insulator falls, at
very low temperatures, into a distinct \textit{zero-conductivity
state}.  A nonmonotonic magnetic-field dependence of $T_0$ and $V_T$
(Fig.~\ref{fig3:RBlow}) naturally results from the magnetic-field
modulation of the effective Josephson energy~\cite{FVB,Vinokur}. The
appealing task now is the construction of the phase diagram of the
zero-conductivity state and revealing the mechanisms of depinning
and temperature crossovers.

\begin{acknowledgments}
We thank D.~Weiss and W.~Wegscheider for access to their high
magnetic field system, V.\,F.~Gantmakher, M.~Feigel'man, A.
Finkelstein, and M. Fistul for useful discussions. This research is
supported by the Program ``Quantum macrophysics'' of the Russian
Academy of Sciences, the Russian Foundation for Basic Research
(Grants No. 06-02-16704 and No. 07-02-00310), the U.S. Department of
Energy Office of Science under the Contract No. DE-AC02-06CH11357,
and the Deutsche Forschungsgemeinschaft within the GRK 638.
\end{acknowledgments}

\end{document}